\documentclass[aps,preprintnumbers,nofootinbib,superscriptaddress,12pt]{revtex4}

\usepackage{graphics}
\usepackage{hyperref}
\usepackage{epsfig}
\usepackage{color}

\usepackage[normalem]{ulem}
\usepackage{multirow}
\usepackage{bm}

\begin{document}

\title{Imprinting spin patterns by local strain control in a van der Waals antiferromagnet} 

\author{Zhuoliang Ni$\dag$}
\affiliation{Department of Physics and Astronomy, University of Pennsylvania, Philadelphia, Pennsylvania 19104, U.S.A.}
\author{Huiqin Zhang$\dag$}
\affiliation{Department of Electrical and Systems Engineering, University of Pennsylvania, Philadelphia, PA, 19104, U.S.A.}
\author{Qi Tian}
\affiliation{Department of Physics and Astronomy, University of Pennsylvania, Philadelphia, Pennsylvania 19104, U.S.A.}
\author{Amanda V. Haglund}
\affiliation{Department of Materials Science and Engineering, University of Tennessee, Knoxville, TN 37996, U.S.A.}
\author{Nan Huang}
\affiliation{Department of Materials Science and Engineering, University of Tennessee, Knoxville, TN 37996, U.S.A.}
\author{Matthew Cothrine}
\affiliation{Department of Materials Science and Engineering, University of Tennessee, Knoxville, TN 37996, U.S.A.}
\author{David G. Mandrus}
\affiliation{Department of Materials Science and Engineering, University of Tennessee, Knoxville, TN 37996, U.S.A.}
\affiliation{Materials Science and Technology Division, Oak Ridge National Laboratory, Oak Ridge, TN, 37831, U.S.A.}
\author{Deep Jariwala$^*$}
\affiliation{Department of Electrical and Systems Engineering, University of Pennsylvania, Philadelphia, PA, 19104, U.S.A.}

\author{Liang Wu$^*$}
\affiliation{Department of Physics and Astronomy, University of Pennsylvania, Philadelphia, Pennsylvania 19104, U.S.A.}

\date{\today}




\maketitle 
\def\thefootnote{$\dag$}\footnotetext{These authors contributed equally to this work}\def\thefootnote{\arabic{footnote}}

\newpage

\textbf{Van der Waals magnets provide opportunities for exploring low-dimensional magnetism and spintronic phenomena. The Mermin-Wagner theorem states that long-range correlations in reduced dimensions are stabilized and controlled by magnetic anisotropy. In this study, we meticulously create and control the inplane easy-axis magnetic anisotropy within two-dimensional (2D) van der Waals antiferromagnet MnPSe$_3$ via a novel method involving topography and therefore strain control by using a micro- patterned substrate. By transposing the MnPSe$_3$ thin flakes onto a substrate patterned with micro-scale grooves, we introduce local uniaxial strain pattern, which not only locks the spin direction to the strain direction but also replicates the groove pattern in the spin orientation distribution.  Our approach generates spin orientations that correspond to the substrate patterns, therefore having the potential to significantly advance spintronic devices by offering a unique method for manipulating and designing spin textures in easy-plane magnets}.

Spintronic devices have garnered significant interest due to their low energy consumption and high operation speeds compared to conventional electronics\cite{chumak2015magnon,baltzrmp2018}. Antiferromagnets, with their advantages of and terahertz  spin waves and the absence of a stray field, are emerging as a new route to control magnetism and engineer memory devices.  Antiferromagnets can have smaller independent devices to increase the storage density compared with the traditional ferromagnetic materials. Two-dimensional antiferromagnets also have the possibility to make memory at the atomically thin limit.  Spin orientation plays a crucial role in the operation of spin devices, including magnon transportation\cite{cornelissen2015long,lebrun2018tunable,XingPRX2019}, magnetic tunnel junctions\cite{yuasa2004giant,parkin2004giant,kleinsci18,wangnatcomm18}, and spin-transfer torque\cite{ralph2008spin}. Therefore, controlling spin orientation in materials is essential for optimizing the functionality and performance of spin-based devices. While previous research has employed magnetic fields, pulsed currents, and uniaxial strains to reorient spins\cite{chumak2015magnon,baltzrmp2018, ninatnano21}, these methods have restricted the capability for spatially variant spin control. 

Van der Waals magnets, featured with low-dimensional magnetism, have gained considerable attention since the discovery of magnetism down to monolayer\cite{huangnat17,gongnat17,leenanoletter16}. Unique phenomena and effects are predicted in 2D magnetic systems because of the reduced dimensionality\cite{burchnat18,gibertininatnano19,maknatrevphys19}. In contrast to 3D magnets, long-range magnetic order is absent without the presence of magnetic anisotropy, as indicated by the Mermin-Wagner theorem\cite{merminprl66}. Indeed, magnetic anisotropy plays a pivotal role: it not only safeguards the magnetic order against thermal perturbations but also assigns a favored orientation for spins. Such characteristics provide 2D magnets a higher degree of tenability. This controllability, combined with a wealth of magnetic phases, presents novel possibilities for spin devices, such as electric gate control\cite{dengnat18,huangnatnano18,wang2018electric}, ultrafast light manipulation\cite{Afanasievsciadv21,matthiesenPRL2023}, and mechanical strain tuning\cite{vsivskinsnatcomm20,wang2020strain,ninatnano21}. Notably, previous research has demonstrated strain-controllable local spin orientations in the van der Waals antiferromagnet MnPSe$_3$ with a N\'eel temperature of 68 K by stretching the samples with a polymer\cite{ninatnano21}. MnPSe$_3$ exhibits XY-like magnetism with in-plane N\'eel order  due to its  easy plane magnetic anisotropy \cite{jeevanandamjpcm1999,calderprb2021}, making the spins highly sensitive to in-plane uniaxial strain\cite{ninatnano21}. The unique properties of MnPSe$_3$ suggest that local spin control can be achieved through the application of designed strain distributions. Previous study indicates the magnetic anisotropy induced by the uniaxial strain is large enough to align the spin orientations along with the uniaxial strain direction, irrespective of the crystal axis\cite{ninatnano21}. Note that a small angle between the uniaxial strain and the spin direction is generally allowed\cite{ninatnano21}. Here are the details for the theory. First, it can be explained by the Landau theory\cite{ninatnano21}. An effective theory of the Landau free energy under the uniaxial strain for the angle of the N\'eel vector direction with respect to the crystal axis, $\theta$,  has the form,
\begin{equation}
F(\theta) =  b u_s L_0^2 \cos 2(\theta - \theta_0) + e L_0^6 \cos 6\theta
\end{equation}
where $L_0$ is the magnitude of the N\'eel vector, and $b$, $e$ are coefficients. $\theta_0$ is one of the principal axes of the uniaxial strain.  For $u_s=0$, this corresponds to zero strain and reduces to the  six ground states due to the three-fold rotational symmetry and the time reversal symmetry, which the   $\cos 6\theta$ term represents. For $u_s \ne 0$ it has the two ground states which have spin directions 180-degree to each other, which the   $\cos 2(\theta-\theta_0)$ term represents. The antiferromagnetic order parameter is locked to the strain axis with a constant tilt angle when minimizing the free energy over $\theta$. A second theory of the spin model on the honeycomb lattice was also developed in Ref.\cite{ninatnano21}, and it gives the same conclusion of the spin-strain locking with an intrinsic angle between the two directions. One can read supplementary notes 5-6 for more details of the theory in the supplementary information of Ref.\cite{ninatnano21}. 

In the present study, we introduce an innovative technique for engineering customized spin mapping by manipulating local strain to create an easy axis in the  easy-plane van der Waals magnet MnPSe$_3$. Our method involves transferring a thin flake of the magnet onto a substrate embossed with a pre-patterned array of grooves. We subsequently map the spin orientation distribution by utilizing an optical second-harmonic generation microscope\cite{fiebigprl94,chuprl20,ninatnano21}. Our observations reveal that the local spin orientation follows the pattern of the underlying grooves. Intriguingly, the spins exhibit a tendency to orient themselves perpendicularly to the groove edges, a behavior attributed to the strain exerted on the samples. Our study contributes a novel method to control the spin orientation distribution in magnetic materials with easy-plane anisotropy.

\begin{figure*}
\centering
\includegraphics[width=\textwidth]{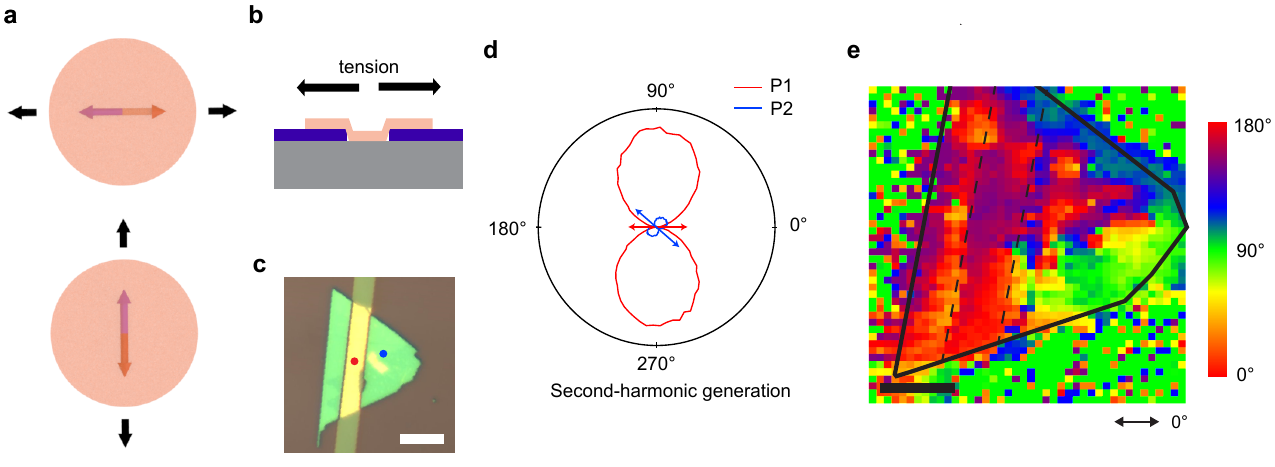}
\caption{\textbf{Spin orientation control on a straight groove.} {\bf a}, Strain-controllable spin orientations of van der Waals XY magnet MnPSe$_3$. The two orange circles represent  MnPSe$_3$ flakes that are not rotated.} The spin orientations are always locked to the uniaxial strain direction, when the applied strain direction is rotated by 90 degrees. {\bf b}, Schematic of the strain direction. {\bf c}, The optical image of the device. A MnPSe$_3$ layer is stacked on a Si/SiO$_2$ substrate with a straight groove. Scale bar: 5 $\mu$m. {\bf d}, Polarization-dependent second-harmonic generation of P1 (on the groove) and P2 (on the substrate). The node direction of the two-fold patterns is the spin orientation of the point, as marked by the arrows. The positions of the two points are market in both {\bf c} and {\bf e}. {\bf e}, The spin mapping across the whole sample. The orientation  of the N\'eel vector is represented by  the color. The black like is the contour of the sample. The  SHG imaging with amplitude information is shown in Extended Data Fig.1a.
\label{Fig1}
\end{figure*}

The MnPSe$_3$ lattice exhibits C$_3$ crystalline symmetry, which implies six equivalently preferred directions of the in-plane spin orientations with the crystalline anisotropy\cite{jeevanandamjpcm1999,Maisciadv2021}. Figure \ref{Fig1} (a) illustrates the relationship between the spins (depicted by red and purple  arrows within the sample) and the strain (indicated by black arrows). In order to induce local strain on the sample, we begin by preparing a SiO$_2$/Si substrate featuring a designed straight groove pattern of etched SiO$_2$, as demonstrated in Figure \ref{Fig1}(b). The SiO$_2$ layer, with a thickness of 90 nm and a width of 4 $\mu$m, is fully etched to achieve the desired pattern. We have tried different groove depths (90 nm, 120 nm, 200 nm) and groove widths (2 $\mu$m, 3 $\mu$m, 4 $\mu$m, 6 $\mu$m). We have found the 90 nm groove thickness and 3-4 $\mu$m groove width makes it best for sample fabrication. Next, we transfer the exfoliated thin-flake sample onto the etched pattern using a polydimethylsiloxane (PDMS) stamping method. The sample thickness is carefully selected to range between 30 nm and 100 nm in order to maintain an optimal balance between the sample's tensile strength and tunability. We have tried some samples above 200 nm, but they are not flexible enough to  attach the samples to the groove bottom. We have also tried some samples around 10 nm, but they are too fragile and easily damaged during the transfer process. Therefore, we think the 30-100 nm samples work the best in our method.

We first investigate the effect of a straight groove on a thin-flake sample, the optical image of which is displayed in Fig. \ref{Fig1} (c). Due to the stretching force exerted by the two edges of the groove, the sample within the groove experiences tensile strain perpendicular to the edge, as illustrated in Fig.\ref{Fig1}(b). To obtain information about the spin orientation, we carry out optical second-harmonic generation (SHG) measurements using an 800 nm wavelength and an objective with a numerical aperture of 0.5, resulting in a spatial resolution of around 2 $\mu$m for the second-harmonic light. The polarization-dependent SHG allows  direct detection of the symmetry information of the sample's second order susceptibility, which is predominantly influenced by the N\'eel vector orientation\cite{ninatnano21}. Figure.\ref{Fig1}(d) presents the SHG results from points located both on the groove (P1) and outside the groove (P2). P1 and P2 are marked with red and blue dots in Fig.\ref{Fig1}c. The polarization dependence is measured by simultaneously rotating the incident polarization and the perpendicular detecting polarization. Based on the symmetry analysis, the nodal direction of the two-fold shape corresponds to the orientation of the N\'eel vector/spin\cite{ninatnano21}. The red and blue arrows indicates the spins on the two Mn sites, which is the same as the N\'eel vector direction. The  N\'eel vector  direction mapping is obtained and depicted in Fig.\ref{Fig1} (e), where the color represents the direction of the N\'eel vector. The complete N\'eel vector  mapping with both direction and amplitude is obtained and depicted in Extended Data Fig.1a, where the color and orientation of each segment represent the spin orientation. The magnitude of the SHG two-fold pattern is denoted by the opacity of the segment at each point. Notably, the majority of the sample within the groove area exhibits spins perpendicular to the groove edges, aligning with our design objective. Conversely, the region outside the groove displays spins generating much smaller SHG magnitudes and random orientations, indicative of a randomly distributed strain. The enhancement of SHG intensity over one order of magnitude inside the groove region can be attributed to the amplification from the interference effects from different interfaces\cite{guo2023ultrathin}, and thus it does not indicate the relationship between the strain and the SHG signal.

\begin{figure*}
\centering
\includegraphics[width=0.8\textwidth]{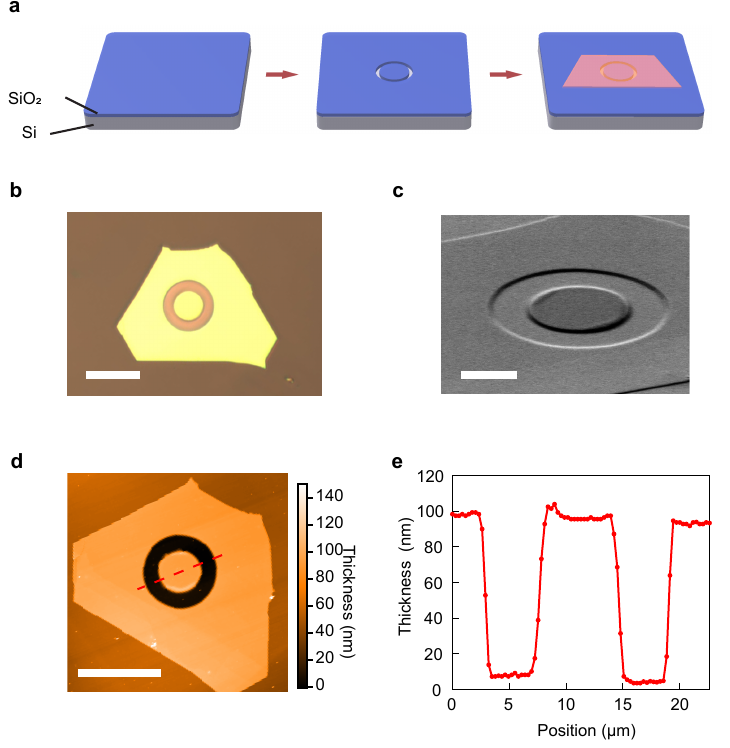}
\caption{\textbf{Spin orientation control on a ring-shaped groove} {\bf a},   Device fabrication. A groove (a circle here) is first etched on a Si/SiO$_2$ substrate (blue). The sample (orange) is then transferred on the groove. The groove would apply strain perpendicular to the edge. {\bf b}, The optical image of the device. Scale bar: 20 $\mu$m. {\bf c}, The scanning electron microscope image of the device. Scale bar: 5 $\mu$m. {\bf d} An atomic force microscopy imaging of the device. Scale bar: 20 $\mu$m. {\bf e} The depth profile along the red dashed line in {\bf d}. (The MnPSe$_3$ sample thickness is 50 nm.) } 
\label{Fig2}
\end{figure*}

To evaluate the flexibility of spin rotation, we place the sample on a ring-shaped groove, as depicted in Fig.\ref{Fig2}(a).  An example of the sample positioned on a ring-shaped groove is characterized through optical microscopy and scanning electron microscopy (SEM), as shown in Figure \ref{Fig2}(b) and (c), respectively. The outer and inner radii of the ring measure 9 $\mu$m and 5 $\mu$m, respectively. The width of the ring groove is 4 $\mu$m. The SEM image clearly reveals that the portion of the sample situated on the groove is securely attached to the Si substrate as one can still see the groove structure clearly.   We also characterize the sample by  atomic force microscopy imaging and show the results in Fig.\ref{Fig2}(d). The sample is quite uniformly attached to the SiO$_2$ and the groove. We take a line-cut measurement along the red-dashed line in Fig.\ref{Fig2}(d), and the thickness step profile is shown in Fig.\ref{Fig2}(e). The majority of the sample within the groove is attached to the groove, indicating that this method to apply a good amount of strain on a ring pattern  is doable.  

\begin{figure*}
\centering
\includegraphics[width=0.8\textwidth]{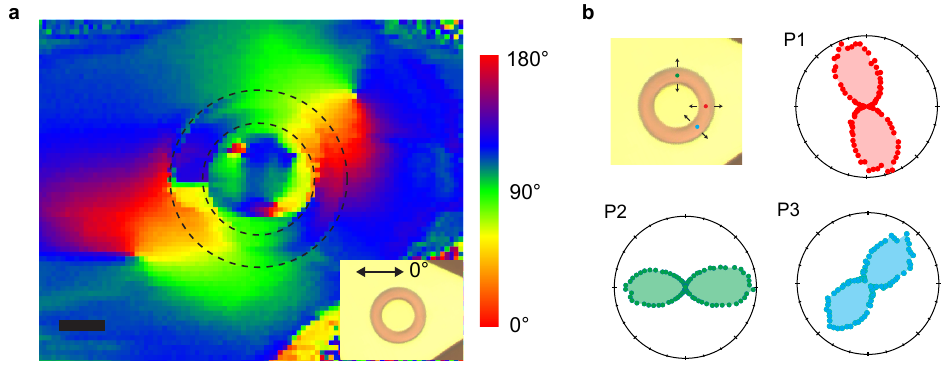}
\caption{\textbf{Imprinting spin patterns on a ring-shape groove.} {\bf a}, The spin pattern imaging of a MnPSe$_3$ flake (thickness: 50 nm) on a ring-shape groove. Scale bar: 5 $\mu$m.} Inset: optical image of the sample. The normalized SHG imaging is shown in Extended Data Fig.1b. {\bf b}, The polarization-resolved SHG measured at different positions on the ring, marked by tiny dots with different colors. The tiny black arrows on the optical image represent the local strain directions. 
\label{Fig3}
\end{figure*}

We also perform SHG imaging on this sample. As shown in Fig. \ref{Fig3}a, the majority of the groove region maintains a spin orientation perpendicular to the tangent of the ring-shaped groove, suggesting that the strain is still applied perpendicularly to the edge of the groove even when the groove is curved. The polarization-dependent SHG patterns of three different points are presented in Fig.\ref{Fig3}(b), clearly demonstrating that the spin orientation on the ring-shaped groove is perpendicular to its tangent. Additionally, we observe that the spin orientation of the region just outside the ring-shaped groove is altered, indicating that the ring-shaped groove applies radial strain both inside and outside the ring region.  The complete N\'eel vector  mapping with both direction and amplitude is obtained and depicted in Extended Data Fig.1b. It is important to note that the optical path in the region outside the groove pattern is closer to the resonant length, resulting in a larger SHG magnitude compared to the region inside the groove.  Due to the laser coherence length is much longer than the sample/SiO$_2$ thickness here, different reflected SHG contributions add up coherently. As a result, the total SHG signal is very sensitive to the situation whether different contributions are in-phase or out-of-phase. We have found these resonant effects are usually dominating the total SHG signal. Therefore, it is not accurate to compare the absolute SHG signal inside and outside the groove without carefully considering the resonant effects. The resonance condition  is not the focus in this paper, but we hope it will generate interests for future studies. 

In conclusion, we have successfully illustrated that substrates featuring irregular patterns can proficiently manipulate spin patterns within van der Waals magnet MnPSe$_3$. Our result showcase the potential for effectively manipulating spin orientations in van der Waals magnets using a variety of groove geometries, further expanding the possibilities for designing spintronic devices. Our investigations uncover that both linear and curved grooves possess the capability to secure the spin orientation in a manner that is perpendicular to the groove edge. This research establishes a foundation for the design of programmable spin distributions, a crucial stride towards advancements in spin devices. Potential applications extend to guided magnon transportation and lateral giant magnetoresistance, thereby unveiling new avenues for the evolution and enhancement of spintronic devices.

\section{Acknowledgments}
L.W. acknowledges support by the US Office of Naval Research (ONR) through the grant N00014-24-1-2064.  The development of the scanning imaging microscope was sponsored by the Army Research Office and was accomplished under Grant Number W911NF-20-2-0166 and W911NF-21-1-0131, and the University Research Foundation.  The sample exfoliation setup is based upon work supported by the Air Force Office of Scientific Research under award number FA9550-22-1-0449. Z.N. acknowledges support from Vagelos Institute of Energy  Science  and  Technology  Graduate  Fellowship and Dissertation Completion Fellowship  at  the  University  of  Pennsylvania. D.G.M acknowledges support from the Gordon and Betty Moore Foundation's EPiQS Initiative, Grant GBMF9069. D.J. acknowledges primary support from Office of Naval Research (ONR) Young Investigator Award (YIP) (N00014-23-1-203) Metamaterials Program. This research was supported in part by grant NSF PHY-1748958 to the Kavli Institute for Theoretical Physics (KITP). A portion of the sample fabrication, assembly and characterization were carried out at the Singh Center for Nanotechnology at the University of Pennsylvania, which is supported by the National Science Foundation (NSF) National Nanotechnology Coordinated Infrastructure Program grant NNCI-1542153.

\section*{Methods}

 \subsection*{Device fabrication}
 Patterning of the ring or straight groove on the SiO$_2$ substrate is achieved by a combination of electron-beam lithography (Elionix ELS-7500EX), exposure of polymethyl methacrylate (PMMA A4) and dry-etching process. Dry-etching is performed using a combination of CHF$_3$ and Ar to fully remove the SiO$_2$. Finally, the samples are cleaned in acetone to remove the PMMA photo-resist and a  ring or straight Si groove is exposed.
 
 Few-layer MnPSe$_3$ flakes are mechanically exfoliated from bulk MnPSe$_3$ crystals using Scotch tape and then dry transferred onto a Si ring or straight groove.

 \subsection*{SHG microscopy}

The SHG microscopy utilizes an 800 nm Ti-sapphire laser (80 MHz, $\sim$ 50 fs) as the fundamental light. The beam is focused by a 50$\times$ objective to the beam spot size of around 1.5 $\mu$m. The same objective is used to collect the reflected 400 nm beam, which is then detected by a photon counter. A half-wave plate and a linear polarizer are used to control the polarization of the incident and outgoing beams, respectively. The SHG image is collected by scanning the sample position on XY piezo stages. All the SHG measurement is performed at 10 K.

\subsection{Author Contribution}
Z.N., and H.Z.  conceived the project. H.Z. and Z.N. made the samples, performed the  measurements and data analysis. A.H., N.H., and D.G. grew the single crystals. Z.N., H.Z., D.J. and L.W. wrote the paper with input from all co-authors. D.J. and L.W. supervised the entire study.

\clearpage
\setcounter{figure}{0}
\renewcommand{\figurename}{{\bf{Extended Data Figure}}}

\begin{figure*}
\centering
\includegraphics[width=0.8\textwidth]{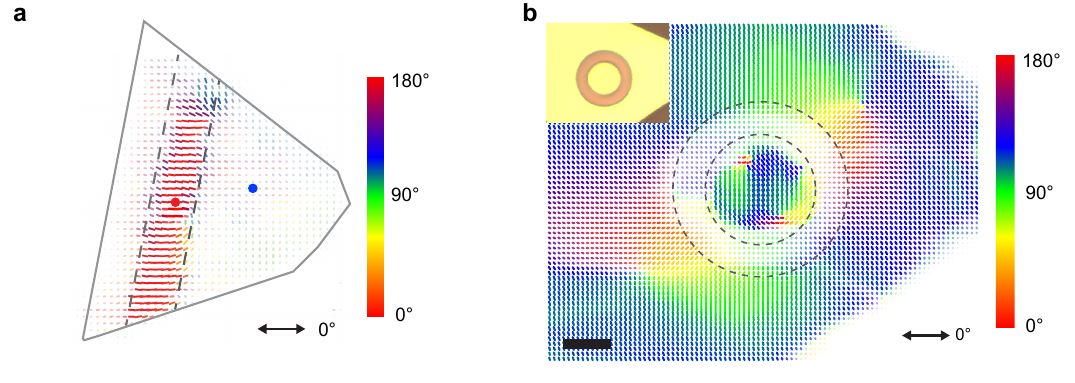}
\caption{\textbf{SHG mapping of the N\'eel vectors of the samples on the (a) straight and (b) circular groove. Scale bar: 5 um. } The orientation and the length of the N\'eel vector are represented by the pointing and the length of the small segment at each point, respectively. To enhance the visualization, the color of the segment is linked to the orientation of the N\'eel vector, and the opacity of the segment is linked to the length of the N\'eel vector.}
\end{figure*}

\section*{Addendum}

\textit{Data availability:} All data needed to evaluate the conclusions in the paper are present in the paper and the Supplementary Information. Additional data related to this paper could be requested from the authors.

\textit{Competing Interests: }The authors declare that they have no
competing financial interests.

\textit{Correspondence: }Correspondence and requests for materials
should be addressed to L.W. (liangwu@sas.upenn.edu) and D.J. (dmj@seas.upenn.edu)


\bibliographystyle{naturemag}
\end{document}